\documentclass{article}
\usepackage{amsmath}
\usepackage{amssymb}
\usepackage{bm}
\usepackage{color}
\usepackage{setspace}
\begin{document}
\begin{center}
\LARGE{Colored Quaternion Dirac Particles\\ of Charges 2/3 and -1/3}
\end{center}
\medskip
\begin{center}
Lester C. Welch

Aiken, SC 29803

lester.welch@gmail.com
\end{center}
\begin{quote}
By starting with the simplest expression of the first order linear wave equation (Dirac's equation) and by confining the elements of the coefficients (matrices) to the quaternions, $\mathbb{H}$, it is shown that structure with three "colors" with charges of 2/3 and -1/3 results with a minimum of assumptions.  It is shown how color neutral particles are required for a positive definite probability.\\ \\
\end{quote}
\newpage

Since the pioneering work of Finkelstein, $\emph{et.,al}$ \cite{Finkelstein2}, quaternions (see Appendix $\mathbf{I}$ for a brief overview of the algebra of quaternions) have received a great deal \cite{Adler,Yefremov,Frenkel,Negi,Leo1,Khalek,Adler1,Maia,Rawat,Rawat2} of attention as a mathematical formulism for expressing physics because of the greater richness of the three non-commutative independent imaginaries ($i,j,k \in \mathbb{H}$).  However in making the transition from the complex field, $\mathbb{C}$,  one must be aware that many theorems, mathematical expressions and formulae\footnote{For a simple example,$e^{ia}e^{ib}= e^{i(a+b)}$, but $e^{ja}e^{ka} \ne e^{(j+k)a}$ } assume that the unit imaginary commutes and thus are not valid.  De Leo and Rotelli \cite{Leo2} also stress this point and caution that variational calculus and tensor analysis are altered from the traditional approach. An important consideration is the one-to-many transition from the commuting imaginary scalar $i=\sqrt{-1}$ to an imaginary in $\mathbb{H}$.  The quaternion ring has received considerable mathematical attention (outside of physics) \cite{Frenkel,Khalek,Feuter,DeLeo,Eilenberg} in an effort to define analyticity, holomorphic functions, and the equivalent to Cauchy's integral formula but apparently no consensus has been reached on the best way to proceed because of the non-commutative nature and the resulting left-right dichotomy.  This article does an \emph{ab initio} development of the Dirac formulism using simpler mathematical techniques in order to avoid the necessity to investigate each theorem and formulism for applicability in non-commuting rings. 
\section{A "Ring-Free" Expression for Dirac's Equation}
In $\mathbb{C}$, Dirac's equation is often given as
\[(i\gamma^\mu\partial_\mu - m)\psi = 0\]
which involves $i \in \mathbb{C}$ and thus forces the first decision point in transitioning to another mathematical ring. 
It is well known (see, for example, \cite{Adler} for a historical review) that only three mathematical rings, $\mathbb{R}$, (the "reals"), $\mathbb{C}$, (the "complexes") and $\mathbb{H}$, (the "quaternions")  are candidates for constructing quantum mechanical descriptions- call them QM$_R$, QM$_C$, and QM$_H$, repectively.  Complexified quaternions, of the form
$q= a+bi+cj+dk;\;\; a,b,c,d \in \mathbb{C}$ have been used in previous works \cite{Yefremov,Negi,Rawat,Rawat2} but are not used in this work as they do not form a division algebra and do not give a meaningful probability \cite{Adler}.  \\ \\ For clarity, to avoid the explicit use of $i$, the most general form ($c = \hbar = 1$) of Dirac's equation is
\begin{equation}
\mathcal{H}\psi=(C_\mu\partial_\mu)\psi=(C_x \partial_x + C_y \partial_y + C_z \partial_z + C_t \partial_t)\psi= m\psi. \label{eqn:dw}
\end{equation}
To recover the Klein-Gordon equation 
\begin{equation}
(\nabla^2 - \partial_t^2)\psi =  m^2\psi , \label{eqn:KG}
\end{equation}
the following conditions must hold\footnote{Both the anticommutator brackets $\{,\}$ and the commutator brackets $[,]$ will be used with their usual definition in this article.} (true for whatever numerical field):
\begin{equation}C_{x,y,z}^2 = 1;\hspace{.1in}  C_t^2 =  -1; \hspace{.1in}\mbox{ and }\left\{C_\mu,C_\nu\right\}=C_\mu C_\nu + C_\nu C_\mu =0, \mbox{ where } \mu\neq\nu.\hspace{.1in}\mu,\nu = x,y,z,t\label{eqn:cnd}
\end{equation}
 \\
Equation ($\ref{eqn:dw}$) can be rewritten\footnote{Throughout this work the position of the indices, $\mu, \nu$ etc have no significance with respect to covariance or contravariance and are placed for typographical convenience.  Repeated indices, however, do indicate summation}
\begin{equation}
(\gamma^\mu\partial_\mu -m)\psi = 0,\hspace{.3in}\mu=0,1,2,3 \label{eqn:qb} 
\end{equation}
by defining
\[\gamma^\mu = (C_t,C_x, C_y, C_z)\] and this avoids the explicit use of an imaginary scalar.  It is understood that the ring-specific scalars will be used to construct the elements of $C_\mu$.

\section{Dirac's Equation in $\mathbb{H}$, QM$_H$}
Adler $\cite{Adler}$ has written an excellent book giving a comprehensive treatment and review of quaternion quantum mechanics, QM$_H$.  He shows that asymptotically QM$_H$ and QM$_C$ give the same result and hopes that QM$_H$ will explain some of the details (e.g., "flavor" and "color") for which it is not clear that QM$_C$ will suffice. This work fits into that scheme in that QM$_H$ is used to investigate these quantum numbers.  It is the philosophy of this work that QM$_C$ - in the guise of quantum electrodynamics - is one of the most successful scientific theories and that QM$_H$ can offer no new insight in that realm.\\ \\  In this work, the ansatz is that QM$_H$ should be interpreted as describing "strongly" interacting fermions - quarks. If quaternion quantum mechanics does not describe quarks - what does it describe?  If quarks - fundamental fermions - are not described by Dirac's equation, what are they described by?  Mathematical elegance and Occam's razor would seem to favor a congruence of the answers of these two questions.  \\ \\
Starting\footnote{ It was shown in $\cite{welch}$, that the results from ($\ref{eqn:dw}$) when using the complex field ($\mathbb{C}$), yield exactly the same formulism as that resulting from using more common representations of Dirac's  gamma matrices- as it should.} with ($\ref{eqn:dw}$), where it is to be understood that the elements of $C_\mu$ and other numbers can now be from the quaternion ring $\mathbb{H}$, to satisfy ($\ref{eqn:cnd}$) four anticommuting quantities are needed. It was shown in \cite{welch} that one choice, of order 4 - the lowest possible, if one avoids\footnote{Yefremov \cite{Yefremov}, De Leo and Rotelli \cite{Leo1} and Rawat and Negi \cite{Rawat,Rawat2} use a 2x2 representation, but Yefremov and Rawat introduces $i \in \mathbb{C}$ and De Leo and Rotelli use a preferred complex plane to form a complex scalar product, and these conditions enable a lower order representation.  Our approach is purely quaternion.}  the use of a fourth imaginary $i \in \mathbb{C}$ throughout the formulism - for the $C_\mu$ is given in Appendix $\mathbf{II}$.  \\ \\
The three complex units (\emph{i,\; j,\; k}) in $\mathbb{H}$ as compared to the single $i \in \mathbb{C}$ means that "complex conjugation" has to be clearly specified.   The following "modes" of conjugation - involutions - are defined:
\[\Gamma_i (q) \overset{def}{=}\; q^{*i} = a - b i + c j + d k,\mbox{     $i$-conjugation}\]
and similarly for $\Gamma_j$ and $\Gamma_k$ - a "single conjugation;"
\[\Gamma_{ij} (q) \overset{def}{=}\;q^{*ij} = a - b i - c j + d k, \mbox{     $ij$ conjugation}\]
and similarly for $\Gamma_{ik}$ and $\Gamma_{jk}$, - a "double" conjugation; and
\[\Gamma_{ijk}(q) \overset{def}{=}\; q^* = a -b i - c j - d k,\mbox{  triple (complete) conjugation, }\]
an operation analogous to "complex conjugation" in $\mathbb{C}$ and is commonly used,  see, for example, \cite{Adler, Leo1} and others.
It is clear that $\Gamma_i, \Gamma_j, \Gamma_k, \Gamma_{ij}$, etc., all commute.  It should also be noted that 
\[q\;\Gamma_{ijk}(q) = qq^* = a^2 - b^2 - c^2 - d^2 = q^*q \in \mathbb{R}\]and
\[\Gamma_{\emph{x}}(rs)=\Gamma_{\emph{x}}(s)\Gamma_{\emph{x}}(r), \mbox{   where}\;\;r,s \in\mathbb{H}\]
\subsection{Single Quaternion Conjugation}
To explore new physics in $\mathbb{H}$, one must look for a symmetry with respect to a particular operator that is available in $\mathbb{H}$ with no parallel in $\mathbb{C}$.  Such operators are the involutions based on conjugations of a subset of $i,\;j,\;k.\;$ It is believed that the first reported use of a conjugation other than a "complete" conjugation was in Welch \cite{welch}. In the following example, $i$-conjugation, will be used but the derivation is the same for $j$-conjugation and $k$-conjugation. 
Suppose
\begin{equation*}
\psi = \left(\begin{array}{c} \psi_1 \\ \psi_2\\ \psi_3 \\ \psi_4\end{array}\right)
\end{equation*}
We define a transpose and $i$-conjugate, of $\psi$
\[\psi^{\dag i} = (\Gamma_i(\psi_1),\Gamma_i(\psi_2),\Gamma_i(\psi_3),\Gamma_i(\psi_4))\]
Taking the transpose and $i$-conjugate   of ($\ref{eqn:qb}$), results in
\begin{equation}
\partial_\mu\psi^{\dag i}(\gamma^\mu)^{\dag i} - m\psi^{\dag i} = 0 \label{eqn:d5}
\end{equation}
Further
\[(\gamma^\mu)^{\dag i} = (\gamma^0, \gamma^1, \gamma^2, \gamma^3)^{\dag i}=  (-\gamma^0, \gamma^1, -\gamma^2, -\gamma^3) \]
one can multiply on the right by $\gamma^3$ and taking advantage of the commutation relationships ($\ref{eqn:cnd}$)
\[(\gamma^\mu)^{\dag i}\gamma^3 =\gamma^3 (\gamma^0,-\gamma^1, \gamma^2, -\gamma^3)\]
and now multiply on the right by $\gamma^2$ to obtain
\[(\gamma^\mu)^{\dag i}\gamma^3\gamma^2 =\gamma^3 \gamma^2(-\gamma^0,\gamma^1, \gamma^2, \gamma^3)\]
and finally multiply by $\gamma^0$ to get
\[(\gamma^\mu)^{\dag i}\gamma^3\gamma^2\gamma^0 =\gamma^3\gamma^2\gamma^0 (-\gamma^0,-\gamma^1, -\gamma^2, -\gamma^3)\]
thus
\[(\gamma^\mu)^{\dag i}\gamma^3\gamma^2\gamma^0 = -\gamma^3\gamma^2\gamma^0\gamma^{\mu}\]
and ($\ref{eqn:d5}$) becomes
\begin{equation*}
\partial_\mu\psi^{\dag i}\gamma^3\gamma^2\gamma^0\gamma^\mu +m\psi^{\dag i}\gamma^3\gamma^2\gamma^0 = 0 
\end{equation*}
and to simplify notation
\[\psi^{\dag i}\gamma^3\gamma^2\gamma^0 = \bar{\psi^i} \] becomes
\begin{equation}
\partial_\mu\bar{\psi^i}\gamma^\mu +m\bar{\psi^i} = 0  \label{eqn:d7}
\end{equation}
Multiplying ($\ref{eqn:d7}$) on the right by $\psi$ and ($\ref{eqn:qb}$) on the left by $\bar{\psi^i}$ and adding gives:
\begin{equation}
\textcolor{red}{\partial_\mu(\bar{\psi^i}\gamma^\mu\psi)  = 0.}
\end{equation}
Likewise
\begin{equation*}
\textcolor{green}{\partial_\mu(\bar{\psi^j}\gamma^\mu\psi)  = 0.}
\end{equation*}
\begin{equation*}
\textcolor{blue}{\partial_\mu(\bar{\psi^k}\gamma^\mu\psi)  = 0.} 
\end{equation*}
\subsection{Double Quaternion Conjugation}
Double quaternion conjugation has the feature that
\[\Gamma_{jk}(\psi) = \Gamma_i(\psi^*)\mbox{,   or}\; \Gamma_{jk}(\psi^*) = \Gamma_i(\psi)\]
so double conjugation in the conjugate space is isomorphic to single conjugation in ordinary space.  It is straight forward to show: ($\bar{\psi}^{jk}=\psi^{\dag jk}\gamma^2\gamma^3 $) 
\begin{equation*}
\textcolor{red}{\partial_\mu(\bar{\psi}^{jk}\gamma^\mu\psi)= 0.} 
\end{equation*}
and likewise:
\begin{equation*}
\textcolor{green}{\partial_\mu(\bar{\psi}^{ki}\gamma^\mu\psi)= 0.}
\end{equation*}
\begin{equation*}
\textcolor{blue}{\partial_\mu(\bar{\psi}^{ij}\gamma^\mu\psi)= 0.}
\end{equation*}

\subsection{Triple (Complete) Quaternion Conjugation}
By taking the transpose and complete conjugation of ($\ref{eqn:qb}$) we get the equation:
\begin{equation}
\partial_\mu\psi^{\dag *}(\gamma^\mu)^{\dag*} - m\psi^{\dag *} = 0 \label{eqn:d2}
\end{equation}
and since
\[(\gamma^\mu)^{\dag*} = (-\gamma^0, \gamma^1, \gamma^2, \gamma^3) \]and by multiplying on the right by $\gamma^0$ gives
\[(\gamma^\mu)^{\dag *}\gamma^0 = \gamma^0(-\gamma^0, -\gamma^1, -\gamma^2, -\gamma^3) =-\gamma^0\gamma^\mu \]
and equation ($\ref{eqn:d2}$) becomes
\begin{equation*}
\partial_\mu\psi^{\dag*}\gamma^0\gamma^\mu +m\psi^{\dag*}\gamma^0 = 0. 
\end{equation*}
By setting
\[\psi^{\dag*}\gamma^0 = \bar{\psi}\] we get
\begin{equation}
\partial_\mu\bar{\psi}\gamma^\mu + m\bar{\psi} = 0 \label{eqn:d3}
\end{equation}
Now by multiplying ($\ref{eqn:qb}$) on the left by $\bar{\psi}$ and ($\ref{eqn:d3}$) on the right by $\psi$ and adding we get
\[\partial_\mu\bar{\psi}\gamma^\mu\psi + \bar{\psi}\gamma^\mu\partial_\mu\psi = 0 \] 
or\begin{equation}\partial_\mu(\bar{\psi}\gamma^\mu\psi) = 0. \label{eqn:c3}\end{equation}
a result analogous to QM$_C$ when using "complex conjugation.".
\section{Quaternion Plane Wave Solutions to Dirac's Equation}
Using the Dirac matrices from Appendix $\mathbf{II}$ in equation($\ref{eqn:qb}$) results in:
\vspace*{.2in}
\begin{equation}
(\mathcal{H}-m) \psi=
\left(\begin{array}{cccc}
-m & i\partial_x +j \partial_y + k\partial_x & -\partial_t & 0\\
-i\partial_x -j \partial_y - k\partial_x &-m&0& \partial_t\\
\partial_t&0&-m&i\partial_x +j \partial_y + k\partial_x\\
0&-\partial_t&-i\partial_x -j \partial_y - k\partial_x&-m
\end{array}
\right)\left(\begin{array}{c} \psi_1 \\ \psi_2 \\ \psi_3 \\ \psi_4 \end{array}\right)=0 \label{eqn:dm}
\end{equation}
\vspace*{.13in}
\newline
It must be noted that if one attempts to make the association of the momentum operator in the x-direction, P$_x$ (say), with $i\partial_x$, then it will not in general commute with the Hamiltonian nor with P$_y$, both of which may be quaternions.  Adler\cite{Adler} highlights this fact and examines the extant efforts to deal with this difficulty and finds none are completely satisfactory.  Adler does show that in the asymptotic region of scattering that P$_x$, etc., have the expected properties, but there is no apparent resolution of this dilemma for the subasymptotic states.\\ \\
Let
$\mathbf{\triangle_q} = i\partial_x +j \partial_y + k\partial_z,\quad(i,j,k \in \mathbb{H})$
thus
\[\mathbf{\triangle_q^2} =- \partial_x^2 - \partial_y^2 -\partial_z^2 = -\mathbf{\triangle^2}\]
and\footnote{As is well known\cite{Adler}, the determinant of a matrix with noncommuting elements does not have the usual properties as matrices with elements $\in\mathbb{C}$.  However in this case all elements commute.} 
\[\mathrm{det}[(\mathcal{H}-m)]=(\mathbf{\triangle^2} - \partial_t^2 -m^2)^2=0\]
and ($\ref{eqn:dm}$) becomes
\begin{eqnarray}
\mathbf{\triangle_q} \psi_2 -\partial_t \psi_3 -m\psi_1 &=& 0 \nonumber \\
\mathbf{\triangle_q} \psi_1 -\partial_t \psi_4 +m\psi_2 &=& 0 \nonumber \\
\mathbf{\triangle_q} \psi_4 +\partial_t \psi_1 -m\psi_3 &=& 0  \nonumber \\
\mathbf{\triangle_q} \psi_3 +\partial_t \psi_2 +m\psi_4 &=& 0  \label{eqn:s1}
\end{eqnarray}
Defining $q_1$ to be a unit imaginary quaternion\footnote{If an arbitrary unit quaternion imaginary is needed in this work then $q_1 = bi+cj+dk$ such that $b^2+c^2+d^2=1$ is used.},
look for plane wave solutions of the form:

\[ \psi_\mu =\phi_\mu e^{q_1(p_x x +p_y y + p_z z -Et)}\]
where $\phi_\mu$ is not a function of $x,y,x,t$ but can be a quaternion, thus in general, $\phi_\mu q \ne q \phi_\mu,$ \newline and $p_x, p_y, p_z,E \in \mathbb{R}$. For the plane wave trial solution,

\[\mathrm{det}[(\mathcal{H}-m)]=(\mathbf{\triangle^2} - \partial_t^2 -m^2)^2\Rightarrow -(p_x^2+p_y^2+p_z^2)+E^2 -m^2.\]
which means that solutions exist only if $E^2=p^2+m^2$, thus $E= \pm \sqrt{p^2+m^2}$. 
Since

\begin{align*}
\mathbf{\triangle_q} \psi_\mu& = (i\partial_x +j \partial_y + k\partial_z)(\phi_\mu e^{q_1(p_x x +p_y y + p_z z -Et)})\\
 &= (i\phi_\mu \partial_x +j\phi_\mu \partial_y +k\phi_\mu\partial_z)(e^{q_1(p_x x +p_y y + p_z z -Et)})\\
 &=(i\phi_\mu q_1 p_x +j\phi_\mu q_1 p_y +k\phi_\mu q_1 p_z)(e^{q_1(p_x x +p_y y + p_z z -Et)})\\
 &=(ip_x +jp_y +kp_z)(\phi_\mu q_1)(e^{q_1(p_x x +p_y y + p_z z -Et)})
\end{align*}
and defining
\[P_q=(ip_x +jp_y +kp_z)\;\;\; \Rightarrow\;\;\; P_q^2 = -(p_x^2 + p_y^2 +p_z^2) \equiv -p^2\] then ($\ref{eqn:s1}$), the wave equations, become
\medskip
\begin{eqnarray*}
P_q\phi_2q_1 e^{q_1(\mathbf{p x}-Et)} +E\phi_3q_1 e^{q_1(\mathbf{p x}-Et)} -m\phi_1 e^{q_1(\mathbf{p x}-Et)}&= & 0, \\
P_q\phi_1q_1 e^{q_1(\mathbf{p x}-Et)} +E\phi_4 q_1 e^{q_1(\mathbf{p x}-Et)} +m\phi_2 e^{q_1(\mathbf{p x}-Et)}&= & 0, \\
P_q\phi_4 q_1e^{q_1(\mathbf{p x}-Et)} -E\phi_1 q_1e^{q_1(\mathbf{p x}-Et)} -m\phi_3 e^{q_1(\mathbf{p x}-Et)}&= & 0, \\
P_q\phi_3 q_1e^{q_1(\mathbf{p x}-Et)} -E\phi_2 q_1 e^{q_1(\mathbf{p x}-Et)} +m\phi_4 e^{q_1(\mathbf{p x}-Et)}&= & 0.
\end{eqnarray*}
or
\begin{eqnarray}
P_q\phi_2  +E\phi_3  +m\phi_1q_1 &=& 0,  \label{eqn:e4}\\
P_q\phi_1  +E\phi_4  -m\phi_2q_1 &=&  0, \nonumber \\
P_q\phi_4  -E\phi_1  +m\phi_3q_1 &=&  0, \nonumber \\
P_q\phi_3  -E\phi_2  -m\phi_4q_1 &=&  0. \nonumber
\end{eqnarray}
Equation ($\ref{eqn:e4}$) has the following 4 independent orthonormal solutions (See Appendix III for the calculation of the normalization factor, which will be supressed from here on)($m \ne 0$):\\ \\
\begin{equation}
\begin{array}{cccc}
u_{\uparrow}=N
\begin{pmatrix}\displaystyle{\frac{P_q q_1}{m}}  \\[.11in]   1  \\[.07in]0  \\[.1in]\displaystyle{\frac{E q_1}{m}} \end{pmatrix};\quad
u_{\downarrow}=N
\begin{pmatrix}\displaystyle{\frac{E q_1}{m}}  \\[.11in] 0 \\[.07in]  1\\[.1in] \displaystyle{\frac{-P_q q_1}{m}} \end{pmatrix}; \quad
d_{\uparrow} =N
\begin{pmatrix}0 \\[.09in] \displaystyle{\frac{-E q_1}{m}}  \\[.14in] \displaystyle{\frac{P_q q_1}{m}} \\[.11in]1 \end{pmatrix};\quad
d_{\downarrow}=N
\begin{pmatrix}1 \\[.11in] \displaystyle{\frac{-P_q q_1}{m}}  \\[.14in]\displaystyle{\frac{-E q_1}{m}} \\[.11in] 0 \end{pmatrix};
\end{array} \label{eqn:fam}
\end{equation}\\ \\
Note that if one simultaneous reverses the sign of q$_1$ and $m$, the solutions remain unchanged. These solutions are valid for both positive and negative energy states and thus these four states are candidates for different spin states of two distinct particles in the same "family" - as indicated. 
If $m=0$, the solutions are:
\begin{equation}
\begin{array}{cccc}
\eta_{u\uparrow}=\frac{1}{E\sqrt{2}}
\begin{pmatrix}0  \\  1  \\  -P_q/E\\  0 \end{pmatrix};\;
\eta_{u\downarrow}=\frac{1}{E\sqrt{2}}
\begin{pmatrix}\ 0  \\ P_q/E  \\   1\\  0\end{pmatrix};\;
\eta_{d\uparrow} =\frac{1}{E\sqrt{2}}
\begin{pmatrix}P_q/E  \\  0  \\  0 \\  1 \end{pmatrix};\;
\eta_{d\downarrow}=\frac{1}{E\sqrt{2}}
\begin{pmatrix}\ 1  \\  0 \\ 0 \\  -P_q/E\end{pmatrix}
\end{array} \label{eqn:eta}
\end{equation}
The complete wave functions are then:
\begin{equation}
\Psi_1 = u_{\uparrow}e^{q_1(\mathbf{p x}-Et)};\;\;\Psi_2 = u_{\downarrow}e^{q_1(\mathbf{p x}-Et)};\;\;
\Psi_3 = d_{\uparrow}e^{q_1(\mathbf{p x}-Et)};\;\;\Psi_4 = d_{\downarrow}e^{q_1(\mathbf{p x}-Et)}. \label{eqn:fam1}
\end{equation}
Then, using the obvious notation
\[
\psi=
\begin{pmatrix} 
\Psi^1\\[.08in] 
\Psi^2
\end{pmatrix}=
\begin{pmatrix} \psi^1 \\  \psi^2 \\  \psi^3 \\ \psi^4 \end{pmatrix} 
\] it is easily shown that
\begin{equation*}
\begin{pmatrix} 0 & 0 & 0 &1 \\ 
                0 & 0& -1& 0 \\ 
                0&1&0&0 \\ 
                -1&0&0&0
\end{pmatrix} 
\begin{pmatrix}
                d_{\uparrow}^1 \\ 
                d_{\uparrow}^2 \\ 
                d_{\uparrow}^3 \\ 
                d_{\uparrow}^4 
\end{pmatrix}
\equiv
\begin{pmatrix} 0 &\sigma \\ 
                 \sigma & 0 
\end{pmatrix}
\begin{pmatrix} 
D_{\uparrow}^1\\ 
D_{\uparrow}^2
\end{pmatrix}
=
\begin{pmatrix}
                D_{\downarrow}^1 \\ 
                D_{\downarrow}^2 
\end{pmatrix}
\;\mbox{and}\;\; 
\begin{pmatrix} 0 & \sigma\\ 
                 \sigma & 0
\end{pmatrix}
\begin{pmatrix} 
D_{\downarrow}^1\\ 
D_{\downarrow}^2
\end{pmatrix}
=
-\begin{pmatrix}
                D_{\uparrow}^1 \\ 
                D_{\uparrow}^2 
\end{pmatrix};
\end{equation*}
and likewise
\begin{equation*}
\begin{pmatrix} 0 & \sigma \\ 
                \sigma& 0 
\end{pmatrix}
\begin{pmatrix} 
U_{\uparrow}^1\\ 
U_{\uparrow}^2
\end{pmatrix}
=
\begin{pmatrix}
                U_{\downarrow}^1 \\ 
                U_{\downarrow}^2
\end{pmatrix};\;\;\;
\begin{pmatrix} 0 & \sigma \\ 
                 \sigma & 0 
\end{pmatrix}
\begin{pmatrix} 
U_{\downarrow}^1\\ 
U_{\downarrow}^2
\end{pmatrix}
=
-\begin{pmatrix}
                U_{\uparrow}^1 \\ 
                U_{\uparrow}^2
\end{pmatrix}\bigskip
\end{equation*}
It is important to note that each of the solutions ($\ref{eqn:fam}$) for $m \ne 0$ has three distinct imaginary components in that $q_1$ is an arbitrary unit imaginary quaternion thus (for example):
\begin{equation}
u_{\uparrow}^r=
\begin{pmatrix}\displaystyle{\frac{ P_q i}{m}}  \\[.11in] 1  \\[.07in] 0 \\[.1in]\displaystyle{\frac{i E}{m}} \end{pmatrix};\qquad
u_{\uparrow}^g=
\begin{pmatrix}\displaystyle{\frac{ P_q j}{m}}  \\[.11in] 1  \\[.07in] 0 \\[.1in]\displaystyle{\frac{j E}{m}} \end{pmatrix};\qquad
u_{\uparrow}^b=
\begin{pmatrix}\displaystyle{\frac{ P_q k}{m}}  \\[.11in] 1  \\[.07in] 0 \\[.1in] \displaystyle{\frac{k E}{m}} \end{pmatrix}. \label{eqn:ceq}
\end{equation}
\\ \\
Thus in general
\[u_{\uparrow}=a_ru_{\uparrow}^r+a_gu_{\uparrow}^g+a_bu_{\uparrow}^b\]
where
$a_r^2+a_g^2+a_b^2=1$ and $a_r, a_g, a_b \in \mathbb{R}$ as explained below.
\section{Multiple Particles}
\subsection{Linear Superposition}
In QM$_C$ if
$H\psi=m\psi$, multiplying $\psi$ by a phase factor, $c=e^{i\theta}$, does not alter the physics: $H(c\psi)=m(c\psi)$ because $[H,c]=0$.  The equivalent, $H(q_1\psi)=m(q_1\psi)$, is not true in QM$_H$, because in general $[H,q] \ne 0$. This difference in commutivity becomes important when considering one of the basic tenets of QM$_C$ - \emph{the principle of superposition of dynamical states} expressed as follows - in $\mathbb{C}$: 
\begin{eqnarray}
&\mbox{if }&H\psi_i=m\psi_i \nonumber \\
&\mbox{then }&H\Psi=m\Psi \nonumber \\ 
&\mbox{where }&\Psi = a_1\psi_1+a_2\psi_2+\cdots\;\;;a_i\in\mathbb{C} \label{eqn:lsp}
\end{eqnarray} 
But in $\mathbb{H}$,
\begin{eqnarray}
&\mbox{if }&H\psi_i=m\psi_i \nonumber \\
&\mbox{then }&H\Psi\ne m\Psi \nonumber \\ 
&\mbox{where }&\Psi = a_1\psi_1+a_2\psi_2+\cdots\;\;;a_i\in\mathbb{H} \label{eqn:1sh}
\end{eqnarray} \\ \\ 
Thus, in $\mathbb{H},\;\; H\Psi = m\Psi$ only if $a_i\in\mathbb{R}$ and therefore the superposition of dynamical states ($\ref{eqn:1sh}$) must be replaced by
\begin{equation*}
\Psi = a_1\psi_1 +a_2\psi_2 +\cdots\;\;;a_i \in \mathbb{R}
\end{equation*} if the principle of linear superposition is to remain valid.
\subsection{Tensor Products}
To form multi-particles from the particles of ($\ref{eqn:fam}$), one needs a meaningful definition of a tensor product $T$ for quaternion arguments, however it is well known \cite{Adler,Finkelstein} that such doesn't exist when using linear superposition forms like ($\ref{eqn:lsp}$) such as $T\langle\psi_1,\psi_2\rangle=\psi_1\psi_2$. The requirement for multilinearity conditions, i.e.,
\begin{align*}
T\langle\psi_1,\psi_2q+\psi_2^{\prime}q^{\prime}\rangle=&T\langle\psi_1,\psi_2\rangle q+T\langle \psi_1,\psi_2^{\prime}\rangle q^{\prime}\\
T\langle\psi_1q+\psi_1^{\prime}q^{\prime},\psi_2\rangle=&T\langle\psi_1,\psi_2\rangle  q+T\langle\psi_1^{\prime},\psi_2\rangle q^{\prime}
\end{align*}
results (see Adler\cite{Adler}, page 245) in additional terms like $[q,\psi]$.
which in QM$_C$, ($q \in \mathbb{C}$) in general is identically equal to zero while in QM$_H$ is identically zero only if $q \in \mathbb{R}.$ 
Thus a meaningful tensor product such as
\[T\langle\psi_1,\psi_2\rangle=\psi_1\psi_2\pm\psi_2\psi_1\]
obeying multilinearity, can be defined in $\mathbb{H}$ \emph{assuming a reasonable extension of the principle of linear superposition}.\\ \\
Finkelstein \emph{et. al.} \cite{Finkelstein} in 1963 (before the concept of quarks or "color" existed) formulated the principle of general Q covariance which states that \emph{it is physically meaningless to compare quaternions at different space-time points except in their intrinsic algebraic properties}.  This principle gives license to consider, in QM$_H$, only the quaternion algebraic property of the wave function.
Using this license the following postulate is made concerning composite particles:\\ \\
\textbf{The Q symmetry postulate:} The quaternion algebraic property, $\mathcal{Q}$, of any composite solution must be invariant under any permutation of $i,j,k$.  The quaternion algebraic property, "quaternion charge", is defined as the sum of the quaternion components of the solutions - ignoring the space-time aspect.  For example if
\begin{eqnarray*}
\Psi_{composite}=T\langle\Psi_n(\mathbf{r_1},q_n),\Psi_m(\mathbf{r_2},q_m)\rangle 
\end{eqnarray*}
then the quaternion charge would be 
\[ \mathcal{Q}=q_n+q_m\]
This concept is useful only to the extent that it yields meaningful results - which it is shown to do in the following.  A similar concept is that of electronic charge where the total charge is the sum of the charges of the individual particles regardless of space-time distribution.

\subsection{Two Quasi-particles}
The Q symmetry postulate means that when a composite of two solutions (particles) of ($\ref{eqn:fam1}$) is made, if one solution has the quaternion charge $q_1$, the other must have $-q_1$ because $\mathcal{Q}=(q_1)+(-q_1)=0$ which is invariant under permutation of $i,j,k$. For example, a composite spin-one particle (choosing $q_1 = i$) could be formed from two solutions of ($\ref{eqn:fam1}$)) as follows:
\begin{eqnarray*}
\Psi_{composite}=T\langle\Psi_{1\uparrow}(\mathbf{r_1},i),\Psi_{3\uparrow}(\mathbf{r_2},-i)\rangle\;\;\mbox{for example;}\\ \\ \rho^+=T\langle u_{\uparrow}(i)e^{i(\mathbf{p_1 r_1}-Et_1)},d_{\uparrow}(-i)e^{-i(\mathbf{p_2 r_2}-Et_2)}\rangle 
\end{eqnarray*}
where
\begin{eqnarray*}
u_{\uparrow}(i)=
\begin{pmatrix}\displaystyle{\frac{P_q i}{m}}  \\[.11in]   1  \\[.07in]0  \\[.1in]\displaystyle{\frac{E i}{m}} \end{pmatrix};\quad
d_{\uparrow}(-i)=
\begin{pmatrix}0,  \\[.11in]  \displaystyle{\frac{(-E) (-i)}{-m}}\\[.11in]\displaystyle{\frac{P_q (-i)}{-m}}    \\[.07in]1  \\[.1in] \end{pmatrix}
\end{eqnarray*}
The signs of \emph{both} $i$ and $m$ have to be reversed in order to continue to be a solution to ($\ref{eqn:e4}$). Hence a particle with an "anti-color" is necessarily an "anti-particle."\\ \\
Likewise, the following has $\mathcal{Q}$ of $0$ and is, thus a possible spin-zero composite particle;
\begin{eqnarray*}
\Psi_{composite}&=T\langle \Psi_{1\uparrow}(\mathbf{r_1},q),\Psi_{2\downarrow}(\mathbf{r_2},-q)\rangle\;\;\mbox{for example;}\\ \\
\pi^0 &= T\langle u_{\uparrow}(q)e^{q(\mathbf{p_1 r_1}-Et_1)},u_{\downarrow}(-q)e^{-q(\mathbf{p_2 r_2}-Et_2)}\rangle 
\end{eqnarray*}
where
\begin{equation*}
u_{\uparrow}(q)=
\begin{pmatrix}\displaystyle{\frac{P_q q}{m}}  \\[.11in]   1  \\[.07in]0  \\[.1in]\displaystyle{\frac{E q}{m}} \end{pmatrix};\quad
u_{\downarrow}(-q)=
\begin{pmatrix}\displaystyle{\frac{E (-q)}{-m}}  \\[.11in] 0 \\[.07in]  1\\[.1in] \displaystyle{\frac{(-P_q)(- q)}{-m}} \end{pmatrix}
\end{equation*}
But the following \textbf{could not} form a composite particle because of the Q symmetry postulate ($\mathcal{Q}=i+k$ is not invariant under the permutation - say - $i \leftrightarrow j$): 
\begin{eqnarray*}
\Psi_{composite}=T\langle\Psi_1(\mathbf{r_1},i),\Psi_1(\mathbf{r_2},k)\rangle= T\langle u_{\uparrow}(i)e^{i(\mathbf{p_1 r_1}-Et_1)},u_{\uparrow}(k)e^{k(\mathbf{p_2 r_2}-Et_2)}\rangle 
\end{eqnarray*}
where
\begin{equation*}
u_{\uparrow}(i)=
\begin{pmatrix}\displaystyle{\frac{P_q i}{m}}  \\[.11in]   1  \\[.07in]0  \\[.1in]\displaystyle{\frac{E i}{m}} \end{pmatrix};\quad
u_{\uparrow}(k)=
\begin{pmatrix}\displaystyle{\frac{P_q k}{m}}  \\[.11in]   1  \\[.07in]0  \\[.1in]\displaystyle{\frac{E k}{m}} \end{pmatrix}
\end{equation*}
\subsection{Three Quasi-particles}
There is only one possible invariant quaternion charge available when forming a three particle composite, namely $\mathcal{Q}=i+j+k$. Thus the only composite possible is of the form
\begin{eqnarray*}
\Psi_{composite}=T\langle\Psi_a(\mathbf{r_1},q_a),\Psi_b(\mathbf{r_2},q_b),\Psi_c(\mathbf{r_3},q_c)\rangle
\end{eqnarray*}
where
$\Psi_a,\Psi_b,\Psi_c$ are solutions of ($\ref{eqn:fam1}$) and $\mathcal{Q}=q_a+q_b+q_c=i+j+k$.
\section{Electrodynamics}
\subsection{Klein-Gordon}
In the presense of an electrodynamic field, $\mathbf{A}$, the Klein-Gordon Equation ($\ref{eqn:KG}$) in $\mathbb{C}$ has the following form (see equation XX.30 of \cite{messiah}):
\begin{equation}
\left[(\partial_r-ieA_r)^2 - (\partial_t-ieA_t)^2\right]\psi =  m^2\psi;\;\;\;r=x,y,z \label{eqn:KG1}
\end{equation}
and it is well known (see \cite{schiff, messiah}) that when an electromagnetic field is present that the solutions to the Dirac equation satisfies a second-order equation that differs from the Klein-Gordon equation by a term coupling the spin to the electromagnetic field. In order to explore any further differences introduced by the use of quaternions a different formulism from what is usually presented will be used to enhance the ability to compare the role of the imaginary bases. We will first consider the Klein-Grodon equation, then the Dirac equation in $\mathbb{C}$ and lastly the Dirac equation in $\mathbb{H}$ to more easily distinguish the difference.
Hence expanding ($\ref{eqn:KG1}$) yields:
\begin{align}
\partial_r^2\psi-\partial_t^2\psi 
- ie\left[\;\partial_r(A_r)- \partial_t(A_t)\;\right]\psi
-2ie\left[\;A_r\partial_r  -A_t\partial_t\;\right]\psi-e^2(A_r^2- A_t^2)\psi&= m^2\psi \label{eqn:kg2} \\ 
\mbox{or,}\;\;\nabla^2\psi -ie \nabla\cdot \mathbf{A} \psi -2ie\mathbf{A}\cdot \nabla \psi -e^2\mathbf{A}^2&=m^2\psi \nonumber
\end{align}
It is this form of the equation, with its explicit manifestation of the imaginary $i\in\mathbb{C}$, which we want for comparison purposes.
\subsection{The Electrodynamic Dirac's Equation in $\mathbb{C}$}
The purpose of this section is to show the consistency of the approach and to create a compatible expression in the current formulism for comparison.  It offers nothing new.  Starting with Dirac's equation ($\ref{eqn:qb}$) and making the minimum substitution yields:
\begin{equation*}
\gamma^\mu(\partial_\mu-G_\mu)\psi= m\psi,\hspace{.3in}\mu=0,1,2,3 \label{eqn:msb} 
\end{equation*}
where the relationship of $G_\mu$ with the electromagnetic field $A_\mu$ will be made later. Applying the left hand operator twice to get a second order equation yields:
\begin{equation*}
\gamma^\mu(\partial_\mu-G_\mu)\gamma^\nu(\partial_\nu-G_\nu)\psi= m^2\psi,\hspace{.3in}\mu,\nu=0,1,2,3 
\end{equation*}
and expanding \\ \\
\begin{equation*}
\gamma^\mu\partial_\mu\gamma^\nu\partial_\nu\psi -\gamma^\mu\partial_\mu\gamma^\nu (G_\nu\psi)-\gamma^\mu G_\mu\gamma^\nu\partial_\nu\psi + \gamma^\mu G_\mu \gamma^\nu G_\nu\psi=m^2\psi 
\end{equation*}
\begin{equation*}
\mbox{or,}\;\;\;\gamma^\mu \gamma^\nu\partial_\mu\partial_\nu\psi -\gamma^\mu\gamma^\nu \partial_\mu(G_\nu)\psi
-\gamma^\mu\gamma^\nu G_\nu\partial_\mu\psi
-\gamma^\mu G_\mu\gamma^\nu\partial_\nu\psi + \gamma^\mu G_\mu \gamma^\nu G_\nu\psi=m^2\psi 
\end{equation*}
where no assumptions have been made about the commutivity among $G$ and $\gamma$.  If $\gamma$ and $G$ do commute, ($\left[\gamma,G\right] = 0$)  - as they do in $\mathbb{C}$ - 
\begin{equation*}
\gamma^\mu \gamma^\nu\partial_\mu\partial_\nu\psi -\gamma^\mu\gamma^\nu \partial_\mu(G_\nu)\psi
-\gamma^\mu\gamma^\nu (G_\nu\partial_\mu
+G_\mu\partial_\nu)\psi + \gamma^\mu \gamma^\nu G_\mu  G_\nu\psi=m^2\psi
\end{equation*}
Setting $G_\mu = ieA_\mu$ and using the commutation relationships ($\ref{eqn:cnd}$) results in:
\begin{equation*}
\nabla^2\psi -ie \nabla\cdot \mathbf{A} \psi -2ie \mathbf{A}\cdot \nabla \psi -e^2\mathbf{A}^2\psi- \frac{ie}{2}\gamma^\mu\gamma^\nu(\partial_\mu A_\nu-\partial_\nu A_\mu)\psi =m^2\psi
\end{equation*}
or defining $\mathbf{S}^{\mu \nu} = \frac{1}{2}\gamma^\mu\gamma^\nu$ and $\mathbf{F}_{\nu\mu} =(\partial_\nu A_\mu-\partial_\mu A_\nu)$ we get\footnote{$\mathbf{S}^{\mu \nu}$ is usually defined to include \emph{i} (see Messiah\cite{messiah}, page 905), but the present definition is adopted to clearly indicate the role of imaginary scalars} 
\begin{align}
\begin{array}{ccc}
\underbrace{\nabla^2\psi -ie \nabla\cdot \mathbf{A} \psi -2ie \mathbf{A}\cdot \nabla \psi -e^2\mathbf{A}^2\psi}\;\;\;+& \underbrace{ie\;\mathbf{S}^{\mu \nu} \mathbf{F}_{\nu\mu}\psi}& =m^2\psi\\ \label{eqn:emc1}
\mbox{        Klein Gordan}\;\;& \mbox{        Spin Orbit}\;\;\;&
\end{array}
\end{align}
\subsection{The Electrodynamic Dirac's Equation in $\mathbb{H}$}
In choosing $G$ from $\mathbb{H}$ a reasonable (and the simplest) choice - analogous to QM$_C$ - is 
\[ G_\mu = e \check{q} A_\mu\]
where  
\begin{equation}
A_\mu(x,y,x,t) \in \mathbb{R},\;\;\; \check{q} \in \mathbb{H},\;\;\check{q}^2 = -1,\;\;\mbox{ and}\;\; \left[\gamma^\mu,\check{q}\right] =0 \label{eqn:qcnd}
\end{equation}
With this choice $\left[\gamma, G\right]=0$ and equation ($\ref{eqn:emc1}$) becomes
\begin{equation}
\nabla^2\psi -\check{q}e \nabla\cdot \mathbf{A} \psi -2\check{q}e \mathbf{A}\cdot \nabla \psi -e^2\mathbf{A}^2\psi+ \check{q}e \;\mathbf{S}^{\mu \nu} \mathbf{F}_{\nu\mu}\psi =m^2\psi \label{eqn:qem}
\end{equation}
The most general form of $\check{q}$, (no scalar $\check{q}$ commutes with all $\gamma^\mu$) imposed by the conditions ($\ref{eqn:qcnd}$) is
\[\check{q} = \left(\begin{array}{cccc} 0&a&b&c\\-a&0&-c&b\\-b&c&0&-a\\-c&-b&a&0\end{array}\right)\] where $a,b,c \in \mathbb{R}\mbox{ and}\;\; a^2+b^2+c^2 =1$, thus there are three linearly independent choices for $\check{q}$, one obvious set (of an infinite number) is:
\[
\check{q}_1 = \left(\begin{array}{cccc} 0&1&0&0\\-1&0&0&0\\0&0&0&-1\\0&0&1&0\end{array}\right)\;\;\;
\check{q}_2 = \left(\begin{array}{cccc} 0&0&1&0\\0&0&0&1\\-1&0&0&0\\0&-1&0&0\end{array}\right)\;\;\;
\check{q}_3 = \left(\begin{array}{cccc} 0&0&0&1\\0&0&-1&0\\0&1&0&0\\-1&0&0&0\end{array}\right)\]
And equation ($\ref{eqn:qem}$) in reality is three different equations:
\begin{equation}
\nabla^2\psi -\check{q}_n e \nabla\cdot \mathbf{A} \psi -2\check{q}_ne \mathbf{A}\cdot \nabla \psi -e^2\mathbf{A}^2\psi+ \check{q}_ne\mathbf{S}^{\mu \nu} \mathbf{F}_{\nu\mu}\psi =m^2\psi\;\;\;\;n=1,2,3 \label{eqn:qmc1}
\end{equation}
So a linear combination, representing a particle composed of three electrically charged quasi-particles, satisfies:
\begin{equation}
(a_1+a_2+a_3)\left[\nabla^2  -e^2\mathbf{A}^2-m^2\right]\psi
-(a_1\check{q}_1+ a_2\check{q}_2+a_3\check{q}_3)\left[ e \nabla\cdot \mathbf{A} \psi +2e \mathbf{A}\cdot \nabla \psi -e\mathbf{S}^{\mu \nu} \mathbf{F}_{\nu\mu}\psi\right] =0 \label{eqn:qmc2}
\end{equation}
If one wishes this equation to be compatible with ($\ref{eqn:emc1}$) when projecting onto a preferred complex plane\footnote{This essentially means that one - in a quaternion equation - surjectively replaces each quaternion imaginary with $i\in\mathbb{C}$ to recover the corresponding complex equation.} then
\[a_1+a_2+a_3=1,\;\;\;\mbox{and}\;\;a_1^2+a_2^2+a_3^2 =1\] To determine $a_1, a_2,$ and $a_3$ one further assumption is needed.  It is reasonable to assume that of the three quasi-particles represented by ($\ref{eqn:qmc2}$), in their lowest energy state, two are identical except for a spin variable.  Pairing of identical fermions of opposite spin is a favored configuration whereever encountered.  Thus we hypothesize that $a_1 = a_2$ and therefore
\[2a_1+a_3=1\;\;\mbox{and}\;\; 2a_1^2 +a_3^2=1\] which has the solution:
\[ a_1 = a_2 = \frac{2}{3}\;\;\mbox{and}\;\; a_3=-\frac{1}{3}\] As a model for the proton, this establishes the charge of the $u$ and the $d$ quark and the proton is composed of 
\begin{eqnarray*}
proton_{\uparrow}&=T\langle u_{\uparrow}(i),u_{\downarrow}(j),d_{\uparrow}(k)\rangle\\
charge&=   2/3 + 2/3 - 1/3 =1
\end{eqnarray*}
Knowing the charges of $u$ and $d$ the other "colorless" composite particles allowed by the postulate of Q symmetry are easily shown to be integer units of $e$.  For example:
\begin{eqnarray*}
neutron_{\uparrow}&=T\langle d_{\uparrow}(i),d_{\downarrow}(j),u_{\uparrow}(k)\rangle\\
charge&=   -1/3   -1/3    +2/3 = 0
\end{eqnarray*}
and for the previous examples:
\begin{eqnarray*}
\rho^+_{\uparrow}&=T\langle u_{\uparrow}(i),d_{\uparrow}(-i)\rangle\\
charge&=   +2/3   +1/3   = 1
\end{eqnarray*}
\begin{eqnarray*}
\pi^0&=T\langle u_{\uparrow}(i),u_{\downarrow}(-i)\rangle\\
charge&=   +2/3   -2/3   = 0
\end{eqnarray*}

\section{Conclusions}
The major results are:
\begin{itemize} 
\item The naturalness of associating "color" with the three imaginary bases of quaternions.
\item Extenting the principle of linear superposition to $\mathbb{H}$ which allows
\item a meaningful tensor product to be formulated  
\item The Q symmetry principle confines the formation of composites to "colorless" configurations, that is, those composites whose algebraic properties do not change under any permutation of the bases, $i,j,k$.  
\item Conserved color currents are shown to exist.  
\item Fractional electronic charges of 2/3 and -1/3 are derived.
\end{itemize}
Attributes of the current scheme:
\begin{itemize} 
\item Only $\mathbb{H}$ is used and thus the formulism meets the criteria required to construct a quantum mechanics.
\item The "spin-orbit" interaction remains unchanged.
\end{itemize}  
Unresolved problems include:
\begin{itemize}
\item The lack of a suitable operator for momentum in the subasymptotic region.
\item The lack of any explanation of "families."
\end{itemize} 
A possible criticism of the approach of this paper is that the successes are due to "numerology" - $\mathbb{H}$ has three imaginaries and there are three colors so it is not surprising that some scheme can be concocted to tie them together. However an counter argument could be, "There are three colors \emph{because} $\mathbb{H}$ is the underlying logic of "strong" quantum mechanics."  The intent of this article is to contribute to the resolution of that debate.
\vspace*{.5in}\\
\noindent
\normalsize{\textbf{Appendix $\mathbf{I}$, Quaternions}} \\ \\
Quaternions, $\mathbb{H}$, are one of only three ($\mathbb{R}$, $\mathbb{C}$ and $\mathbb{H}$) finite-dimensional division rings containing the real numbers  $\mathbb{R}$ as a subring - a requirement to preserve probability in quantum mechanics. $\mathbb{H}$ can be loosely viewed as a non-commutative extension of $\mathbb{C}$.  The imaginary quaternion units, i, j, k are defined by
\[ii = jj = kk = -1\] 
\[ij = -ji = k,\hspace{.3in}ki = -ik = j,\hspace{.3in}jk = - kj = i\]
A general quaternion \emph{q} can be written
\[ q = a + bi + cj + dk\]
where \[\emph{a,b,c,d} \in \mathbb{R}.\]
Every non-zero quaternion has an inverse.  
Quaternion addition is associative  - $q_1 + (q_2 + q_3) = (q_1 + q_2) + q_3$ - and defined as 
\[q_1 + q_2 = a_1 + a_2 +(b_1 + b_2)i + (c_1 +c_2)j + (d_1 +d_2)k\]
and quaternion multiplication (paying heed to the non-commutative nature of the imaginary units) is
\[q_1q_2 = (a_1a_2 - b_1b_2 - c_1c_2 - d_1d_2) \]
\[+(a_1b_2 + b_1a_2 + c_1d_2 - d_1c_2)i \]
\[+(a_1c_2 - b_1d_2 + c_1a_2 + d_1b_2)j \]
\[+(a_1d_2 + b_1c_2 - c_1b_2 + d_1a_2)k \]
Quaternions are associative under multiplication  $(q_1q_2)q_3 = q_1(q_2q_3)$.
A unit imaginary quaternion $\bar{q}$ is defined as
\[\bar{q} = b i + c j + d k\]
\begin{center} ( b, c, d $\in$ $\mathbb{R}$ )
\end{center}
where $\bar{q}^2 = -1$, which means $b^2 + c^2 + d^2 = 1.$
It should also be noted that if \emph{i} refers to the \emph{i} of $\mathbb{C}$ rather than of $\mathbb{H}$ it will be specifically indicated.
\vspace{.5in}\\
\noindent\normalsize{\textbf{Appendix $\mathbf{II}$, Dirac's Matrices in $\mathbb{H}$}}
\vspace*{.1in}
\newline
One representation, maintaining $\emph{\;i,\;j,\;k\;}$ symmetry, satisfying ($\ref{eqn:dw}$) is 
\vspace*{.1in}

$\gamma^0\equiv \gamma^t \equiv C_t =
\left( \begin{array}{cccccccc}
0 & 0 & -1 & 0 \\
0 & 0 & 0 & 1 \\
1 & 0 & 0 & 0\\
0 & -1 & 0 & 0 
\end{array} \right)
\equiv
\left( \begin{array}{cc}
      0& -\mathbf{\sigma}_t  \\
      \mathbf{\sigma}_t& 0
      \end{array}\right);\hspace{.2in}
        \sigma_t=
   \left( \begin{array}{cc}
      1 & 0\\
      0 & -1
          \end{array}\right).$ 
\vspace*{.1in}

$\gamma^1 \equiv \gamma^x \equiv C_x =
\left( \begin{array}{cccc}
 0 & i & 0 & 0\\
 -i & 0 & 0 & 0\\
 0 & 0 & 0 & i\\
 0 & 0 & -i & 0
      \end{array} \right) 
\equiv
i\left( \begin{array}{cc}
      \mathbf{\sigma} & 0 \\
      0 &\mathbf{\sigma}
      \end{array}\right);\hspace{.2in}
        \sigma=
   \left( \begin{array}{cc}
      0 & 1\\
      -1 & 0
          \end{array}\right)$
\vspace*{.1in}

$\gamma^2 \equiv \gamma^y \equiv C_y =
\left( \begin{array}{cccc}
0 & j & 0 & 0\\
-j & 0 & 0 & 0\\
0 & 0 & 0 & j\\
0 & 0 & -j & 0
      \end{array} \right)
\equiv
j\left( \begin{array}{cc}
      \mathbf{\sigma} & 0 \\
      0 &\mathbf{\sigma}
      \end{array}\right)$             
\vspace*{.3in}

$\gamma^3\equiv \gamma^z \equiv C_z = 
\left( \begin{array}{cccccccc}
0 & k & 0 & 0\\
-k & 0 & 0 & 0\\
0 & 0 & 0 & k\\
0 & 0 & -k & 0
       \end{array} \right)
 \equiv
k\left( \begin{array}{cc}
      \mathbf{\sigma} & 0 \\
      0 &\mathbf{\sigma}
      \end{array}\right)$
\vspace*{.5in} \\ \\
\noindent{\normalsize\textbf{Appendix $\mathbf{III}$ Orthonormality}
\vspace*{.15in}
\normalsize
\newline
We define the $\emph{norm}, \;\;\mathbb{N},\;$ as the symmetric form (valid also for QM$_C$)
\[\mathbb{N}(q_1,q_2)=\mathbb{N}(q_2,q_1)=\frac{1}{2}(q_1^\dag q_2 + q_2^\dag q_1)\;\;\;\Rightarrow \mathbb{N}(q) \equiv \mathbb{N}(q,q)=q^\dag q\] 
Using $u_{\uparrow}$ as an example, which is (a reminder from equation($\ref{eqn:ceq}$))
\begin{equation}
u_{\uparrow}=N
\begin{pmatrix}\displaystyle{\frac{ P_q i}{m}}  \\[.11in] 1  \\[.07in] 0 \\[.1in]\displaystyle{\frac{i E}{m}}\end{pmatrix}
\end{equation}
the normalizing factor, $N$, is calculated from
\begin{equation*}
\mathbb{N}(u_{\uparrow})=N^2\begin{pmatrix}\displaystyle{\frac{(-i)(- P_q) }{m}}& 1&0&\displaystyle{\frac{-i E}{m}}\end{pmatrix}
\begin{pmatrix}\displaystyle{\frac{ P_q i}{m}}  \\[.07in] 1  \\[.07in] 0 \\[.07in]\displaystyle{\frac{i E}{m}}\end{pmatrix} = \frac{N^2}{m^2}\;[i(P_q)(P_q)i + m^2 +E^2]=\frac{N^2}{m^2}\;[p^2+m^2+E^2] = 2N^2\frac{E^2}{m^2}
\end{equation*}
thus 
\[N= \left|\frac{1}{\sqrt{2}}\frac{m}{E}\right|\]\\ \\
The wave functions $u_{\uparrow},\;u_{\downarrow},\;d_{\uparrow},\; u_{\downarrow}$ are easily shown to be mutually orthogonal. 
\bibliography{q2rev}
\bibliographystyle{unsrt}
\end{document}